\documentstyle[12pt]{article}

\font\Bf = cmbx12 scaled \magstep 1
\begin{document}
\hsize= 6 truein
\vsize =9 truein
\rightline{\bf UCVFC-DF/5-96}
\rightline{\bf CAT/96-03}
\rightline{\bf hep-th/9606120}
\hrule
\vskip 2cm
\begin{center}
{\Bf Topological sectors and gauge invariance 
in massive vector-tensor theories in D$\geq$4 }
\end{center}

\,

\begin{center}
{\bf P\'{\i}o J. Arias$^{a,b,}$\footnote{email: parias@tierra.ciens.ucv.ve} 
and  Lorenzo Leal$^{a,}$\footnote{email: lleal@tierra.ciens.ucv.ve}}
\end{center}

\,

\begin{center}
${}^a${\it
Grupo de Campos y Part\'{\i}culas\\
Departamento de F\'{\i}sica, Facultad de Ciencias\\
Universidad Central de Venezuela, AP 47270, Caracas 1041-A\\
Venezuela}\\
\vskip 0.5 truecm

${}^b${\it Centro de Astrof\'{\i}sica Te\'orica, Facultad de Ciencias,\\
Universidad de Los Andes,\\ 
AP 26, La Hechicera, M\'erida 5101,\\ 
Venezuela}
\end{center}

\,

\rule{0cm}{0.5cm}

\begin{abstract}
\noindent 
A family of locally equivalent models is considered. They can be taken as a 
generalization to $d+1$ dimensions of the Topological Massive and 
``Self-dual'' models in 2+1 dimensions. The corresponding 3+1 models 
are analized in detail. It is shown that one model can be 
seen as a gauge fixed version of the other, and their space of classical 
solutions differs in a topological sector represented by the classical 
solutions of a pure BF model. The topological sector can be gauged out 
on cohomologically trivial base manifolds but on general settings it may be 
responsible of the difference in the long distance behaviour of the 
models. The presence of this 
topological sector appears explicitly in the partition function of the 
theories. The generalization of this models to higher dimensions is shown 
to be straightfoward.
\end{abstract}

\newpage
One of the motivations for studying field theories in 2+1 dimensions is that, 
being more tractable, one hopes to get some insight on their higher 
dimensional generalizations. This picture becomes more interesting when the 
lower dimensional models provide new ideas for the higher dimensional ones. 
This is the case of the so called ``string'' fractional statistics model
~\cite{rodolfo}~\cite{bala}, which constitutes a generalization of the 
fractional statistics concept in 2+1 dimensions~\cite{frac}. 
In the former example the role of the topological Chern-Simons term in 2+1 
dimensions is generalized by an, also topological, BF term. In both cases the 
statistics appears as a manifestation of the topological structure of the base 
manifold.

The non-trivial topological nature of the base manifold may impose conditions 
on the equivalence between different physical models. In these situations, the 
possible global contributions of the topological terms to the observables of 
the theories may restrict their relation to hold on cohomological trivial 
sectors of the base manifold. This is the scheme 
between two different descriptions of massive spin 1 excitations in 2+1 
dimensions: the ``Self-dual'' (SD)~\cite{sd} and the Topological Massive (TM) 
models~\cite{schon}~\cite{djt}. On simply connected manifolds these two models are 
completely equivalent~\cite{dj}, and it can be shown that the SD model correspond 
to a gauge fixed version of the TM gauge theory~\cite{grs}. Nevertheless, the 
space of solutions of both theories could be different. In fact, beside their  
common solutions there is a topological sector in the space of solutions of 
the TM model not present in the SD one. This topological sector is filled by 
all the flat connections on the base manifold~\cite{ar}. This will not 
constitute any obstacle on simply connected manifolds, because this flat 
connections could be gauged out in the TM model. But on general settings, the 
gauge fixing procedure can only be performed locally, so the equivalence 
between both models will be conditioned to this level. This situation of 
global inequivalence persists 
if we use the usual Stuckelberg form of the SD model. Instead, to get a  
global relation between both models, we have to modify the 
SD action adding to the potential $a_{\mu}$ a closed but not necessarily exact 
1-form $\omega_{\mu}$~\cite{ar}. So, the global equivalence is obtained 
patching and sewing ``SD formulations'' over simply connected sectors of the 
base manifold. The so obtained modified SD action is gauge invariant and 
corresponds to a pure Chern-Simons model superposed on the original SD one
~\cite{ar2}. As it could be expected, on simply connected sectors, the 
modified SD action turns to be the Stuckelberg form of the original one. It 
can also be shown, in a path integral approach, that the TM model can be 
obtained as a dualized version of the SD one~\cite{stephany}.

In this letter we will show that this scheme of local and global equivalence 
between the SD and TM models, and their gauge fixing relation, can be 
generalized to higher space-time dimensions. We first study the 
generalization to 3+1 dimensions. The two models to be considered are 
well known and their comparision with the 2+1 picture has been noticed and 
used in different contexts~\cite{lahiri}\cite{bos}. It will be shown that one 
of the models can be 
taken locally as a gauge fixed version of the other. Also we will prove that 
on base manifolds, with a non-trivial topological structure, both models
might have 
different long-distance behaviour. This difference, as in the 2+1 analogs, 
is due to a topological sector in the space of classical solutions which is 
not common between both models. This topological sector corresponds in $d$+1  
dimensions to 
the classical solutions of a BF model. The presence of this sector is shown to 
appear in the partition function of the gauge invariant model. The 
generalization to $d$+1 dimensions is straihgtfoward through the formulation 
of both models in terms of the duals of the antisymmetric tensors.

In 3+1 dimensions massive spin 1 excitations can be described by the gauge 
invariant action~\cite{cremer}
\begin{equation}
S_{TM}^4 = \int d^4x\Bigl[-{1 \over 4}
F_{\mu\nu}F^{\mu\nu}-{1 \over 12\mu^2}H_{\mu\nu\lambda}H^{\mu\nu\lambda}
-{1 \over 4}\varepsilon^{\mu\nu\lambda\rho}B_{\mu\nu}F_{\lambda\rho}
\Bigl]\;, 
\end{equation}
where $H_{\mu\nu\lambda}=\partial_{\mu}B_{\nu\lambda}+\partial_{\lambda}B_
{\mu\nu}+ \partial_{\nu}B_{\lambda\mu}$ and $F_{\mu\nu}=\partial_{\mu}A_{\nu}- 
\partial_{\nu}A_{\mu}$ represent, respectively, the Kalb-Ramond and Maxwell  
field strengths. $S_{TM}^4$ is invariant (up to a total divergence) under 
the gauge transformations $\delta B_{\mu\nu}=\partial_{\mu}\xi_{\nu}-
\partial_{\nu}\xi_{\mu}$, $\delta A_{\mu}=\partial_{\mu}\lambda$ and constitutes  
a generalization, to 3+1 dimensions, of the TM model~\cite{lahiri}. In this 
model the two polarization states of the Maxwell field combine with the 
unique degree of freedom of the Kalb-Ramond field to produce a massive spin 1 
excitation~\cite{cremer}$-$\cite{lahiri}. The equations of motion 
that arise from $S_{TM}^4$ are
\begin{eqnarray}
\partial_{\nu}F^{\nu\mu}-{1 \over 2}\varepsilon^{\mu\nu\lambda\rho}
\partial_{\nu}B_{\lambda\rho}&=&0~, 
\label{2} \\
{1 \over \mu^2}\partial_{\lambda}H^{\lambda\mu\nu}-\varepsilon^{\mu\nu\lambda
\rho}\partial_{\lambda}A_{\rho}&=&0 \;, 
\label{3}
\end{eqnarray}
where we notice that closed forms $A=A_{\mu}dx^{\mu}$ and 
$B=B_{\mu\nu}dx^{\mu}\wedge dx^{\nu}$ (with $dA=0$ and $dB=0$) 
are always solutions of the 
system. The relation with the Proca theory is obtained by direct inspection: 
from its  equation of motion, $\partial_{\nu}F^{\nu\mu}-\mu^{2}A^{\mu}=0$, 
we see that $A_{\mu}$ is transverse (or it is a co-closed 1-form), so it 
can be thought locally as the dual of an exact 3-form (or a co-exact 2-form); 
this is the second term in (\ref{2}) and equation (\ref{3}) ensures the 
identification. In other direction, the non-abelian generalization of this 
model, proposed by Freedman and Townsend~\cite{ft}, can be obtained from 
$S_{TM}^4$ using the 
self-interaction mechanism~\cite{adel}.

The local relation between $S_{TM}^4$ and the Proca model justify the 
comparison with the first order form of the latter~\cite{ft}~\cite{town} 
\begin{equation}
S_{P}^4 = \int d^4x\Bigl[{1 \over 4}\varepsilon^{\mu\nu\lambda\rho}B_{\mu\nu}
F_{\lambda\rho} -{1 \over 4}B_{\mu\nu}B^{\mu\nu}-{\mu^2 \over 2}A_{\mu}A^{\mu}
\Bigl]\;, 
\end{equation}
which is, also, a first order form of the massive Kalb-Ramond model
~\cite{lahiri}$-$\cite{town}, albeit this model has a 
``spin jump'' in the zero mass limit~\cite{op}~\cite{hagen}~\cite{aurilia}
~\cite{dts}~\cite{town}.

The equations of motion of $S_{P}^4$ are
\begin{eqnarray}
{1 \over 2}\varepsilon^{\mu\nu\lambda\rho}\partial_{\nu}B_{\lambda\rho}
-\mu^{2}A_{\mu}&=&0~, \\
\varepsilon^{\mu\nu\lambda\rho}\partial_{\lambda}A_{\rho}-B^{\mu\nu}&=&0 \;, 
\end{eqnarray}
where we observe that non-zero closed forms $A$ and $B$ do not belong to the 
space of solutions. So on general manifolds there would be a topological 
sector in the space of solutions of $S_{TM}^4$ not present in the 
corresponding space of the model described by $S_{P}^4$. We recognize in 
$S_{P}^4$ the generalization, to 3+1 dimensions, of the SD model. 

The above mentioned models can be rewritten as
\begin{equation}
{S_{P}^4}^{\star}=\int d^4x\Bigl[{1 \over 2}T^{\mu\nu}F_{\mu\nu}
+{1 \over 4}T^{\mu\nu}T_{\mu\nu}-{{\mu}^2 \over 2}A^{\mu}A_{\mu}\Bigl]
\; ,
\label{Pd}
\end{equation}
and
\begin{equation}
{S_{TM}^4}^{\star}=\int d^4x\Bigl[{1 \over 2\mu^2}\partial_{\mu}T^{\mu\nu}
\partial_{\lambda}T^{\lambda}{}_{\nu}-{1 \over 4}F_{\mu\nu}F^{\mu\nu}-
{1 \over 2}T^{\mu\nu}F_{\mu\nu}\Bigl] \;,
\label{TMd}
\end{equation}
where $T^{\mu\nu}\equiv{1 \over 2}{\varepsilon}^{\mu\nu\lambda\rho}
B_{\lambda\rho}$ are the components of ${}^{\star}{B}$. ${S_{TM}}^{\star}$ 
is invariant under the gauge transformations 
$\delta A_{\mu}=\partial_{\mu}\lambda$ and 
$\delta T^{\mu\nu}=\varepsilon^{\mu\nu\lambda\rho}\partial_{\lambda}
{\xi}_{\rho}$. 
The topological sector is now filled by closed 1-forms $A$ and co-closed 
2-forms $T$, which are allways solutions of ${S_{TM}}^{\star}$. 
The generalization, to $d$+1 dimensions of $S_{P}^4$ and 
$S_{TM}^{4}$ is obtain directly from (\ref{Pd}) and (\ref{TMd}) if we use the 
identification $T={}^{\star}B$ with B a ($d-1$)-form. We will keep then 
working in 3+1 dimensions and the results to higher dimensions are trivially 
generalized taking care of the identification.

Let us start showing the canonical equivalence of $S_{P}^4$   
and $S_{TM}^4$ over a cohomological trivial region of space-time. 
We suppose that the base manifold is $M_4=R\times{\Sigma}_3$, with 
${\Sigma}_3$ a compact orientable 3-manifold. Starting with $S_{TM}^4$,   
after performing the canonical analysis we arrive to the hamiltonian density
\begin{eqnarray}
{\cal H}_{TM}^4=\mu^2\Pi_{ij}\Pi_{ij}+{1 \over 2}\Pi_i\Pi_i&+&{1 \over 4}B_{ij}
B_{ij}+{1 \over 2}\varepsilon_{ijk}\Pi_iB_{jk}+ \nonumber \\
&+&{1 \over 4}F_{ij}F_{ij}+{1 \over {12\mu^2}}H_{ijk}H_{ijk} \;,
\end{eqnarray}
subject to the first class constraints $\Theta_a$
\begin{eqnarray}
\theta&=&-\partial_i\Pi_i \qquad \qquad \qquad~, \\
\theta_i&=&-\partial_j\Pi_{ji}+{1 \over 2}\varepsilon_{ijk}\partial_jA_k \;,
\end{eqnarray}
where $\Pi_i$ and $\Pi_{ij}$ are the conjugated momenta associated to $A_i$ 
and $B_{ij}$ (our metric signature is ($-+++$)). The non-canonical variables 
$A_0$ and $B_{0i}$ appear as Lagrange multipliers associated to the 
constraints $\Theta_a$. This set of constraints is reducible (because 
$\partial_i\theta_i=0$) and implies 
the residual gauge invariance $\delta B_{0i}=-\partial_i\xi$.

Going to $S_{P}^4$, after eliminating $A_0$ and $B_{0i}$, we will arrive, 
taking the kinetic part as $\dot{B}_{ij}\varepsilon_{ijk}A_k$, 
to the hamiltonian density
\begin{eqnarray}
{\cal H}_{P}^4={{\mu^2} \over 2}A_iA_i+{1 \over 4}B_{ij}B_{ij}&+&{1 \over 4}
F_{ij}F_{ij}+ \nonumber \\
&+&{1 \over {12\mu^2}}H_{ijk}H_{ijk} \;,
\end{eqnarray}
and the second class constraints $\Phi_A$
\begin{eqnarray}
\varphi_i&=&\Pi_i \qquad \qquad \qquad \qquad \quad , \label{A} \\
\varphi_{ij}&=&\Pi_{ij}+{1 \over 2}\varepsilon_{ijk}A_k\equiv 
\varepsilon_{ijk}\Psi_k \;.
\label{B}
\end{eqnarray}
\noindent The algebra of the constraints $\varphi_i, \Psi_k$ has the only 
non-vanishing equal time Poisson brackets 
$\bigl\{\varphi_i(x),\Psi_j(y)\bigr\}=-(1/2)\delta_{ij}\delta^3(x-y)$. 
This allows us to take half of the 
constraints in (\ref{A},\ref{B}) as first class 
constraints $\Theta_a$, and the other half as gauge fixing conditions
$\Upsilon_a$~\cite{grs}~\cite{biz}. We take $\Theta_a=(-\partial_i\varphi_i, 
\varepsilon_{ijk}\partial_j\Psi_k)=(\theta, {\theta}_i^T)$, $\Upsilon_a=
(-\partial_i\Psi_i, \varepsilon_{ijk}\partial_j\varphi_k)=(\Upsilon, {\Upsilon}
_i^T)$. The bi-directional identification of the sets 
$\Phi_A\leftrightarrow(\Theta_a,\Upsilon_a)\equiv\phi_A$ is possible only 
on sectors where 
the first and second cohomology groups in ${\Sigma}_3$ are trivial, 
so the harmonic parts are taken to be zero. This division in first and second 
class constraints incite us to think on the underlying gauge theory. 
So we look for the gauge invariant hamiltonian~\cite{grs}
\begin{eqnarray}
\widetilde{H}_P^4={H}_P^4&+&\int d^3x \Bigl[\Bigl[ \alpha_a(x)\Theta_a(x)+ 
\beta_a(x)\Upsilon_a(x)\Bigr]+ \nonumber \\
&+&\int d^3y\Bigl[\beta_{AB}(x,y)\phi_A(x)\phi_B(y)\Bigr]\bigr] 
\qquad \quad \;,
\end{eqnarray}
which differs from ${H}_P^4$ by combinations of the constraints, and satisfies 
homogeneous Poisson brackets with the defined first class constraints. Some 
of the coefficients, like the $\alpha$'s, will remain arbitrary. But there is  
a particular solution for wich we get $\widetilde{H}_P^4={H}_{TM}^4$. 
This relation can be written explicitly as
\begin{eqnarray}
\widetilde{H}_P^4&=&{H}_P^4+\int d^3x\Bigl[{1 \over 2}\varphi_i
(\varphi_i+\varepsilon
_{ijk}B_{jk})+2{\mu}^2\Psi_i(\Psi_i-A_i)\Bigr] \nonumber \\
&\equiv&{H}_{TM}^4 \qquad \qquad \qquad \qquad \qquad 
\qquad \qquad \qquad \qquad \;.
\label{16}
\end{eqnarray}
\noindent If we go to the functional integral (the partition function), 
the measure~\cite{sen} takes the form
\begin{equation}
det{\bigl\{\Phi_A,\Phi_B\bigr\}}^{1 \over 2}\delta(\Phi_A)=
det{\bigl\{\Theta_a,\Upsilon_b\bigr\}}\delta(\theta)\delta({\theta}_i^T)   
\delta(\Upsilon)\delta({\Upsilon}_i^T)
\;,
\label{medida} 
\end{equation}
and it can be shown that the right-hand side of this equation is the measure 
we would get in the functional integral of $S_{TM}^4$ after reducing it to the 
independent physical modes~\cite{am}. In fact, in the process to obtain the  
effective, BRST invariant, action of $S_{TM}^4$ we find that due to the 
reducibility property of $\theta_i$ there is a residual gauge invariance 
that must be fixed. This residual invariance comes from the  
arbitrariness in the longitudinal parts of not only $B_{0i}$, 
as we said, but 
also of the pair of ghost-antighost ($D_i$, ${\overline D}_i$) accompanying  
$\theta_i$ and the Lagrange multiplier ($E_i$) associated with the gauge 
fixing constraint~\cite{am}. 
In order to fix these residual invariances in a BRST invariant way we must 
introduce triplets (ghost, antighost, multiplier) for each invariance. Let the 
triplet due to $B_{0i}$ be ($d,{\overline d},b$) and the triplets due to $D_i$, 
${\overline D}_i$ and $E_i$ be respectively ($d_{\overline a}, 
{\overline d}_{\overline a}, b_{\overline a}$), with ${\overline a}=1,2,3$. 
The non-null BRST transformation 
of the ghosts are ($\delta_{BRST}F=\zeta{\hat \delta}F$ with $\delta_{ 
BRST}^2F=0$)
\begin{equation}
\begin{array}{cccc}
{\hat \delta}D_i=\partial_id_1 & 
{\hat \delta}{\overline D}_i=-E_i+\partial_id_2 & 
{\hat \delta}E_i=\partial_id_3 & 
{\hat \delta}{\overline d}_{\overline a}=-b_{\overline a} \\
{\hat \delta}d_2=d_3 &
{\hat \delta}d={\dot d}_1 & 
{\hat \delta}{\overline d}=-b & \\
\end{array}
\;.
\end{equation}
For $A_{\mu}$ and $B_{\mu\nu}$, the transformations are
\begin{equation}
{\hat \delta}A_{\mu}=\partial_{\mu}C,\quad {\hat \delta}B_{ij}=\partial_iD_j-
\partial_jD_i,\quad {\hat \delta}B_{0i}={\dot D}_i-\partial_id  \;,
\end{equation}
where $C$ is the ghost of the triplet ($C$, ${\overline C}$, $E$) associated 
to the gauge invariance of $A_i$. The parity of the involved fields is clear 
from the context if we take account that ${\hat \delta}$ changes it. A good 
gauge fixing condition of these 
residual invariances results to be the cancellation of the projection 
of $B_{0i}$, $D_i$, ${\overline D}_i$ and $E_i$ 
in its longitudinal parts, {\it i.e.}
\begin{equation}
\Upsilon_D=\partial_iD_i,\qquad 
\Upsilon_{\overline D}=\partial_i{\overline D}_i, \qquad 
\Upsilon_E=\partial_iE_i,\qquad \Upsilon_B=\partial_iB_{0i} \;.
\end{equation}
\noindent The effective lagrangian will be~\cite{am} $\sim p\dot{q}
-{\cal H}_{TM}-A_0\theta-B_{0i}\theta_i+\hat{\delta}({\overline D}_{\cal A}
\Upsilon_{\cal A})$, where ${\overline D}_{\cal A}$ and $\Upsilon_{\cal A}$ 
stands, respectively, for the antighosts (of {\it all} the triplets) 
that where introduced, and the corresponding 
gauge fixing conditions. $p$ and $q$ abreviate $\Pi_i$, $\Pi_{ij}$ and $A_i$, 
$B_{ij}$, respectively. 
Now having all the gauge freedom fixed we go 
to the functional integral and start its reduction to the genuine 
physical modes. For this, we integrate all the ``ghosts for ghosts'' and the 
additionally introduced multipliers, arriving to
\begin{equation}
{Z_{TM}}^{red}=\int{\cal D}\Gamma\rho e^{i\int{\cal L}} \;,
\end{equation}
with 
\begin{equation} 
{\cal L}\sim p\dot{q}-{\cal H}_{TM}-A_0\theta-B_{0i}\theta_i-E\Upsilon-
E_i\Upsilon_i+\int d^3y{\overline D}_a\bigl\{\Upsilon_a,\Theta_b(y)\bigl\}D_b(y) 
\;,
\end{equation}
${\cal D}\Gamma={\cal D}p{\cal D}q{\cal D}D_a{\cal D}{\overline D}_a
{\cal D}E_a{\cal D}A_0{\cal D}B_{0i}$, and 
\begin{equation}
\rho=\delta(D_{(i)}^L)\delta({\overline D}_{(i)}^L)\delta(B_{(0i)}^L)
\delta(E_{(i)}^L) 
\nonumber \;.
\end{equation}
\noindent Also, $\Upsilon_a$ are the gauge fixing conditions defined before. 
Integrating the remaining fields excepting the $p$'s and 
$q$'s we arrive to
\begin{equation}
Z_{TM}^{red}=\int{\cal D}p{\cal D}q
det{\bigl\{\Theta_a,\Upsilon_b\bigr\}}\delta(\theta)\delta({\theta}_i^T)   
\delta(\Upsilon)\delta({\Upsilon}_i^T)e^{i\int(p\dot{q}-{\cal H}_{TM})}
\;,
\end{equation}
where we see that the measure in the path integral corresponds to the 
right-hand side of (\ref{medida}), as we asserted. Following with 
(\ref{medida}) and taking care of (\ref{16})
\begin{eqnarray}
Z_{TM}^{red}&=&\int{\cal D}p{\cal D}q
det{\bigl\{\Phi_A,\Phi_B\bigr\}}^{1 \over 2}\delta(\Phi_A)
e^{i\int(p\dot{q}-{\cal H}_{TM})} \nonumber \\
&=&\int{\cal D}A_{\mu}{\cal D}B_{\mu\nu}e^{iS_P^4} \qquad \qquad \qquad \qquad 
\;. 
\end{eqnarray}
\noindent Then, on cohomological trivial 
sectors of the base manifold the covariant effective action of ${S}_{TM}^4$ 
will be ${S}_P^4$, stating that under this condition 
the latter action can be seen as a gauge 
fixed version of the former. On general grounds to have a global canonical 
equivalence, we have to modify ${S}_P^4$ in order to include the topological 
sectors originally absent in its space of solutions. This inclusion will modify 
the partition function by a factor that represents the mentioned sectors. 
These and other feature can be elucidated considering the master 
action
\begin{eqnarray}
S_{M}^{4} =&&\int d^4x\Bigl[-{1 \over 4}
b_{\mu\nu}b^{\mu\nu}-{\mu^2 \over 2}a_{\mu}a^{\mu} 
+{1 \over 3!}\varepsilon^{\mu\nu\lambda\rho}
a_{\mu}H_{\nu\lambda\rho}+ \nonumber \\
&&\qquad \qquad +{1 \over 4}\varepsilon^{\mu\nu\lambda\rho}
(b_{\mu\nu}-B_{\mu\nu})F_{\lambda\rho}\Bigl] 
\qquad \qquad \qquad \;. 
\end{eqnarray}
\noindent This action has the same gauge invariances of $S_{TM}^{4}$ 
(with $b_{\mu\nu}$ and $a_{\mu}$ transforming homogenously). Its dual field  
version is
\begin{eqnarray}
{S_{M}^{4}}^{\star} =&&\int d^4x\Bigl[{1 \over 4}
t_{\mu\nu}t^{\mu\nu}-{\mu^2 \over 2}a_{\mu}a^{\mu} 
+{1 \over 2}(t^{\mu\nu}-T^{\mu\nu})F_{\mu\nu}+ \nonumber \\
&&\qquad \qquad +{1 \over 2}a_{\mu}\partial_{\nu}T^{\nu\mu}
\Bigl] 
\qquad \qquad \qquad \;,
\label{Md}
\end{eqnarray}
where $T={}^{\star}B$, as before, and $t={}^{\star}b$.

From $S_{M}^4$ we obtain the equations of motion
\begin{eqnarray}
b^{\mu\nu}&=&
{1 \over 2}\varepsilon^{\mu\nu\lambda\rho}F_{\lambda\rho} \qquad \qquad , 
\label{8} \\
a^{\mu}&=&{1 \over \mu^{2}3!}\varepsilon^{\mu\nu\lambda\rho}
H_{\nu\lambda\rho}, 
\label{9} \\
\varepsilon^{\mu\nu\lambda\rho}
\partial_{\lambda}(A_{\rho}-a_{\rho})&=&0
\qquad \qquad \qquad  , 
\label{10}\\
\varepsilon^{\mu\nu\lambda\rho}
\partial_{\nu}(B_{\lambda\rho}-b_{\lambda\rho})&=&0 
 \qquad \qquad \qquad  \;.
\label{11}
\end{eqnarray}
\noindent Using (\ref{8}) and (\ref{9}) in $S_{M}^{4}$, the second order 
action $S_{TM}^{4}$ is obtained. By the other side, from (\ref{10}) we 
learn that $a_{\mu}$ and $A_{\mu}$ differ by 
a closed form $\omega_{\mu}$. Also, using (\ref{11}), an analogous situation 
occurs between $B_{\mu\nu}$ and 
$b_{\mu\nu}$ (let the corresponding closed 
form be $\Omega_{\mu\nu}$). Locally we can set 
$\omega_{\mu}=\partial_{\mu}\lambda$ and 
$\Omega_{\mu\nu}=\partial_{\mu}L_{\nu}-\partial_{\nu}L_{\mu}
\equiv{\cal G}_{\mu\nu}$ and going now into $S_{M}^{4}$ 
we obtain a Stuckelberg form of $S_{P}^{4}$
\begin{eqnarray}
S_{St} =&& \int d^4x\Bigl[{1 \over 4}
\varepsilon^{\mu\nu\lambda\rho}
B_{\mu\nu}F_{\lambda\rho}-{\mu^2 \over 2}(A_{\mu}-\partial_{\mu}\lambda)
(A^{\mu}-\partial^{\mu}\lambda)\nonumber \\ 
&& \qquad \qquad-{1 \over 4}(B_{\mu\nu}-{\cal G}_{\mu\nu})
(B^{\mu\nu}-{\cal G}^{\mu\nu})\Bigl] 
\qquad \qquad \qquad\;, 
\end{eqnarray}
which is invariant under $\delta A_{\mu}=\partial_{\mu}\xi$, 
$\delta B_{\mu\nu}=\partial_{\mu}{\xi}_{\nu}-\partial_{\nu}{\xi}_{\mu}$, 
$\delta \lambda=\xi$, $\delta L_{\mu}={\xi}_{\mu}+\partial_{\mu}\chi$. 
The exact forms can be gauged out and we recover $S_{P}^{4}$, 
stating the local equivalence between the models. 

In general the solutions of (\ref{10}) and (\ref{11}) are as 
we stated:  $a_{\mu}=A_{\mu}-\omega_{\mu}$ and 
$b_{\mu\nu}=B_{\mu\nu}-\Omega_{\mu\nu}$. 
This mantains the homogenity of $a_{\mu}$  
and $b_{\mu\nu}$ under gauge transformations. Going to 
${S}_{M}^{4}$, we will obtain the gauge invariant action
\begin{eqnarray}
\widetilde{S}_{P}^{4} =&&\int d^4x\Bigl[
-{1 \over 4}\varepsilon^{\mu\nu\lambda\rho}
\Omega_{\mu\nu}F_{\lambda\rho}+
{1 \over 3!}\varepsilon^{\mu\nu\lambda\rho}(A_{\mu}-\omega_{\mu})
H_{\nu\lambda\rho}- \qquad \qquad \qquad \nonumber \\ 
&&\qquad \qquad -{1 \over 4}(B_{\mu\nu}-\Omega_{\mu\nu})
(B^{\mu\nu}-\Omega^{\mu\nu})
-{\mu^2 \over 2}(A_{\mu}-\omega_{\mu})(A^{\mu}-\omega^{\mu})
\Bigl]\;. 
\label{ZZ}
\end{eqnarray}
\noindent The latter action is global and locally equivalent to $S_{TM}^{4}$, 
and it has incorporated the topological sectors not present, originally, in 
$S_{P}^{4}$. One important feature 
of $\widetilde{S}_{P}^{4}$ is that $\omega_{\mu}$ and 
$\Omega_{\mu\nu}$ can be taken as independent fields and they will be closed 
forms dynamically. So $\widetilde{S}_{P}^{4}$ is the correct modification to 
$S_{P}^{4}$ in order to obtain a complete 
correspondence with $S_{TM}^{4}$. The gauge invariances of 
$\widetilde{S}_{P}^{4}$ are the ones on $S_{TM}^{4}$ plus 
$\delta{\omega}_{\mu}=\delta A_{\mu}$, 
$\delta{\Omega}_{\mu\nu}=\delta B_{\mu\nu}$. In a different but equivalent 
approach we can eliminate $A_{\mu}$ and $B_{\mu\nu}$ 
in $S_{M}^{4}$ with (\ref{10}) and (\ref{11}) (in this case 
$A_{\mu}=a_{\mu}+\omega_{\mu}$, 
$B_{\mu\nu}=b_{\mu\nu}+\Omega_{\mu\nu}$). Doing so, we arrive to the 
pair of uncoupled actions 
\begin{eqnarray}
\widetilde{S}_{P}^{4}\Bigl[a,b,\omega,\Omega\Bigl]
&=&S_{P}^{4}\bigl[f,h\bigl]-{1 \over 2}
\int d^4x\varepsilon^{\mu\nu\lambda\rho}
\Omega_{\mu\nu}\partial_{\lambda}\omega_{\rho} \nonumber \\
&\equiv&S_{P}^{4}\bigl[a,b\bigl]-S_{BF}^{4}\bigl[
\omega_1,\Omega_{2}\bigl] \;,
\label{dec}
\end{eqnarray}
where $S_{BF}^{4}$ is the part that describes the topological sectors 
incorporated only in $S_{TM}^{4}$, and $S_{P}^{4}$ describes the local 
physical degrees of freedom. Taking into account the substitution we just 
made and equation (\ref{dec}) we notice that $A_{\mu}=A_{\mu}^{P}+A_{\mu}^{BF}$, and 
$B_{\mu\nu}=B_{\mu\nu}^{P}+B_{\mu\nu}^{BF}$ belong to the space of solutions 
of $S_{TM}^{4}$ (this assertion holds even in presence of external sources). 
The space of gauge inequivalent classical solutions of the BF theory, when 
the base manifold is $M_{4}=R\times{\Sigma}_3$, is a direct sum of the 
first and second de Rham cohomology groups on ${\Sigma}_3$, and by 
Hodge's duality this space is even dimensional~\cite{hor}. Because of the topological  
character of the BF theory it will not contribute to the physical spectrum
but the long distance behaviour of the solutions of ${S}_{TM}^{4}$, when the 
field strengths tend to zero asymptotically, will be 
characterized by the periods of the BF's solutions while all this periods 
cancel, in this limit, for the Proca theory. Let us illustrate this fact 
considering $S_P^4$ and $S_{TM}^4$ in presence of a point charge 
($J^0=e\delta^3(\vec{x})$, $J^{i}=0$) and a vortex ($J^{0i}={g \over 2}\oint_C 
dy^i\delta^3(\vec{x}-\vec{y})$,$ J^{ij}=0$). The exterior static solutions are
\begin{eqnarray}
A_0^{TM}&=&A_0^P=-eY(\vec{x}),\\
A_i^{TM}&=&A_i^{P}+A_i^{BF}=\Bigl[-g\varepsilon_{ijk}\partial_j\oint_Cdy^k
Y(\vec{x}-\vec{y})\Bigl]+ \nonumber \\
& &\qquad \qquad \qquad \qquad \qquad 
+\Bigl[g\varepsilon_{ijk}\partial_j\oint_Cdy^kC(\vec{x}-\vec{y})+
\partial_i\lambda\Bigl], \\
B_{0i}^{TM}&=&{B_{0i}}^{P}+B_{0i}^{BF}= -\mu^2g\oint_Cdy^iY(\vec{x}-\vec{y})+
\partial_iB,\\
B_{ij}^{TM}&=&B_{ij}^{P}+B_{ij}^{BF}=
\Bigl[e\varepsilon_{ijk}\partial_kY(\vec{x})\Bigl]
+\Bigl[-e\varepsilon_{ijk}\partial_kC(\vec{x})+
\partial_ib_j^t-\partial_jb_i^t\Bigl]\;,
\end{eqnarray}
where $C(\vec{x})=\bigl[4\pi\vert\vec{x}\vert\bigl]^{-1}$ and 
$Y(\vec{x})=\bigl[4\pi\vert\vec{x}\vert\bigl]^{-1}
e^{-\mu^2\vert\vec{x}\vert}$ 
are respectively the Coulomb and Yukawa Green functions 
($(-\Delta+\mu^2)Y(\vec{x})=(-\Delta)C(\vec{x})=\delta^3(\vec{x})$), 
and the arbitrariness in $\lambda$, $B$ and $b_i^t$ ($\partial_ib_i^t=0$) 
due to 
gauge invariance is shown. These solutions are well defined outside sources 
and in this region $H_{\mu\nu\lambda}^{BF}=0$, $F_{\mu\nu}^{BF}=0$, while for 
the Proca solutions the field strengths tend to zero asymptotically. 
If we take an sphere of radius R surrounding the 
origin we get
\begin{equation}
I_B^{BF} \equiv\oint_{\vert\vec{x}\vert=R}B_{ij}^{BF}dx^i \wedge dx^j=2e\;.
\end{equation}
This value is independent of the closed surface and is zero when the charge is outside. 
For the Proca solution
\begin{equation}
I_B^P=-2e(1+\mu R)e^{-\mu R} \;,
\end{equation}
and we note that $I_B^P \to 0$ as $R \to +\infty$, so $I_B^{TM} \to I_B^{BF}$ 
in this limit. $I_B^{BF}$ is the period of the closed 2-form 
$B=B_{ij}dx^i\wedge dx^j$ and we see, as we stated, that this period labels 
the TM solutions asymptotically.

For $A_i$, we have
\begin{eqnarray}
I_A^{BF}&=&\oint_{C'}dx^iA_i=g\varepsilon_{ijk}\oint_{C'}dx^i\oint_{C}dy^j
{(x^k-y^k) \over {\vert\vec{x}-\vec{y}\vert}^3} \nonumber \\
&=&-{g \over 4\pi}\int ds\int ds'({\partial{\hat u} \over \partial s}\times
{\partial{\hat u} \over \partial s'})\cdot{\hat u} \nonumber \\
&=&-gL(C',C) \;,
\end {eqnarray}
where $L(C',C)$ is the linking number of the closed paths $C'$ and $C$. 
The unit vector ${\hat u}(s,s')$ is defined by the parametrization of the 
paths as ${\hat u}=\vert\vec{R}(s,s')\vert^{-1}\vec{R}(s,s')$, with 
$\vec{R}(s,s')=\vec{x}(s')-\vec{y}(s)$. $I_A^{BF}$ corresponds to the period 
of the closed 1-form $A=A_idx^i$, and it is a topological invariant. For the 
Proca solution we will get 
\begin{equation}
I_A^P={g \over 4\pi}\int ds\int ds'({\partial{\hat u} \over \partial s}\times
{\partial{\hat u} \over \partial s'})\cdot{\hat u}(1+\mu R(s,s'))
e^{-\mu R(s,s')}\;.
\end{equation}
\noindent This integral is not a topological invariant and becomes 
negligible when the paths are, point to point, far apart. So, also in this 
aspect the TM  and Proca solutions have different behaviour.

Now, to end our discussion of the 3+1 models we note that a path integral 
approach tells us, from (\ref{ZZ}), that the
partition function $\widetilde{Z}_{P}^{4}$ is equal to $Z_{TM}^{4}$, up to a 
factor 
independent of the fields. This is obtained integrating the ``omegas''. 
From (\ref{dec}) we obtain that the partition 
function of ${S}_{TM}^{4}$ and ${S}_{P}^{4}$ differ by a topological 
factor 
\begin{equation}
Z_{TM}^{4}=Z_{BF}^{4}Z_{P}^{4}\;. 
\end{equation}
This topological factor, $Z_{BF}^{4}$, 
is proportional to the Ray-Singer analytic torsion of the manifold $M_4$
~\cite{tor}~\cite{top}~\cite{hor}. To see this we perform the canonical 
analysis of 
$S_{BF}^4$ and note that the ghost for ghost structure is analogous to 
the one in $S^{TM}$. To obtain the covariant effective action we make the 
identifications: 
\begin{equation}
D_{\mu}=(d,D_i), {\overline D}_{\mu}=({\overline d}, 
{\overline D}_i), E_{\mu}=({\dot d}-b, E_i)\;, 
\end{equation}
so ${\hat \delta}D_{\mu}=\partial_{\mu}d_1$, 
${\hat \delta}{\overline D}_{\mu}=-E_{\mu}+\partial_{\mu}d_2$, 
${\hat \delta}E_{\mu}=\partial_{\mu}d_3$ and 
${\hat \delta}B_{\mu\nu}=\partial_{\mu}D_{\mu}-\partial_{\nu}D_{\mu}$. In 
the covariant Lorentz gauge, the BRST invariant effective action results to be
\begin{equation}
S_{eff}^{BF}=S_{\cal B}+S_{\cal F}\;,
\end{equation}
where the bosonic part is

\begin {eqnarray}
S_{\cal B}=\int d^4 x\Bigl[{1\over 4}
\varepsilon^{\mu\nu\lambda\rho}B_{\mu\nu}F_{\lambda\rho}&-&
E\partial^{\mu}A_{\mu}-E^{\mu}\partial^{\nu}B_{\mu\nu}
-b_3\partial^{\mu}E_{\mu} \nonumber \\
&+&{\overline d}_1\partial_{\mu}\partial_{\mu}d_1+{\overline d}_2(
\partial^{\mu}\partial_{\mu}d_2-\partial^{\mu}E_{\mu})\Bigl] \;,
\end{eqnarray}
and the fermionic part is
\begin{eqnarray}
S_{\cal F}=-\int d^4 x\Bigl[b_1\partial_{\mu}D_{\mu}&+&b_2\partial_{\mu}
{\overline D}_{\mu}+{\overline C}\partial_{\mu}\partial_{\mu}C \nonumber \\
&+&{\overline d}_3\partial_{\mu}\partial_{\mu}d_3+
{\overline D}_{\mu}\partial_{\nu}(\partial^{\mu}D^{\nu}-
\partial^{\nu}D_{\mu})\Bigl]  \;.
\end{eqnarray}

Now, we take $S_{eff}$ on a compact 
Riemmanian manifold $M_4$ without boundary where we have the inner product 
between p-forms $(\omega_p\vert\gamma_p)=\int_{M_4}\omega_p\wedge 
\star\gamma_p$ so the adjoint exterior derivative is 
$\delta_p=(-1)^{np+n+1}\star d \star$. The Laplacian on p-forms is, as usual, 
$\Delta_p=\delta_{p-1}d+d\delta_p$. On $M_4$, $S_{\cal B}$ and $S_{\cal F}$,  
take the form
\begin{eqnarray}
S_{\cal B}=&&{1 \over 2}(B\vert\star dA)-(b\vert\delta A)-
{1 \over 2}(E\vert\delta B)-(b_3\vert\delta E) \nonumber \\
&&+({\overline d}_1\vert\Delta_0d_1)+({\overline d}_2\vert\Delta_0d_2-\delta E) 
\\
S_{\cal F}=&&-(b_1\vert\delta D)-(b_2\vert\delta{\overline D})
-({\overline C}\vert\Delta_0C)-({\overline d}_3\vert\Delta_0d_3)+
({\overline D}\vert\delta dD)\;,
\end{eqnarray}
where $D=D_{\mu}dx^{\mu}$, ${\overline D}={\overline D}_{\mu}dx^{\mu}$, 
$E=E_{\mu}dx^{\mu}$. Integrating the bosonic fields in the path integral 
we will get $Z_{\cal B}=
(det\Delta_0)^{-{5 \over 2}}(det\Delta_1)^{-{1 \over 2}}(det\Delta_2)
^{-{1 \over 4}}$, up to a field independent factor. 
Doing first the b's integration in the fermionic part, 
and then the others we obtain $Z_{\cal F}=(det\Delta_0)^2det\Delta_1$, up to 
an, also, field independent factor. We must 
observe that up to this point we have assumed the absence of zero 
modes. This does not contradict our previous arguments because the path 
integration is made over the coexact pieces of all the fields involved, with  
their exact pieces gauged fixed. Then
\begin{equation}
Z_{BF}^{4}=\int\bigl[{\cal{D}}h\bigl]T^{-{1 \over 4}}(M_4)\;,
\label{aaa}
\end{equation}
where $\bigl[{\cal{D}}h\bigl]$ indicates that an integration 
over the zero-modes remains to be done, and $T(M_4)$ represents the Ray-Singer 
analitical torsion of $M_4$
\begin{equation}
T(M_4)=(det\Delta_0)^2(det\Delta_1)^{-2}det\Delta_2 \;,
\label{bbb}
\end{equation}
with the determinants computed {\it via} $\zeta$-function 
regularization~\cite{tor}, so only non-zero eigenvalues contribute. When 
the manifold is cohomologically trivial (so there are no zero-modes) 
$det\Delta_2=(det\Delta_1)^2(det\Delta_0)^{-2}$ 
(Proposition 4 in~\cite{top}), then $T(M_4)=1$ and $Z_{BF}^{4}=1$, ensuring 
the complete equivalence between $S_{TM}^{4}$ and $S_{P}^{4}$. In general, 
the zero mode integration will give a factor that is also a topological 
invariant. For an even dimensional compact manifold without boundary the Ray-Singer torsion 
is trivial, but fron (\ref{aaa}) we observe that $Z_{BF}\neq1$. The 
integration over the zero modes must be kept in order to have an appropiate 
path integral measure for computing expectation values~\cite{STy}~\cite{hor}. 
This integration runs over a graded sum of cohomology groups due to the 
alternating parity of the fields involved~\cite{STy}~\cite{top}. 
For odd dimensional compact manifolds without boundary the Ray-Singer is in 
general non-trivial, even in the absence of zero modes.

Finally, we quote that all these results are generalized trivially to $d$+1 
dimensions. The corresponding models are written as (\ref{Pd}) and (\ref{TMd}) 
or equivalently in terms of the ($d-1$)-form $B$
\begin{equation}
S_{P}^{d+1}=\int_{M_{d+1}}\Bigl[{1\over 2}B_{d-1} \wedge F+
{1 \over 8}B_{d-1} \wedge{}^{\star}B_{d-1}+
{\mu^2 \over 2}A \wedge{}^{\star}A\Bigl] \;,
\label{p}
\end{equation}
and
\begin{equation}
S_{TM}^{d+1}=\int_{M_{d+1}}\Bigl[{1 \over 8\mu^2}H\wedge{}^{\star}H+
{1 \over 2}F \wedge{}^{\star}F-{1 \over 2}B_{d-1} \wedge F \Bigl]
\;,
\label{tm}
\end{equation}
where $H=dB$ and $F=dA$. These actions can be extended to $d$=2. In the latter 
case each one of the corresponding models describe two massive spin 1 
excitations as the 
Proca model in 2+1 dimensions. For $d\geq$3 the connection between (\ref{p}) 
and (\ref{tm}) is analogous 
to that of the 3+1 analized models: 
\begin{itemize}
\item Both models describe the same physical spectrum as the Proca  
model, which is described by $d$ independent physical degrees of 
freedom.
\item $S_{P}^{d+1}$ is locally a gauge fixed version of 
$S_{TM}^{d+1}$ .
\item $S_{TM}^{d+1}$ has a topological sector in its space of 
solutions not present in the former. This topological sector corresponds to 
the space of classical 
solutions of the BF model (with Lagrangian density ${\cal L}_{BF}=B \wedge dA$), 
and is responsible of the different 
long distance behaviour of the physical models, where the field strengths tend 
to zero asymptotically. 
\item The presence of the 
topological sector appears as a topological factor in their partition 
functions: $Z_{TM}^{d+1}=Z_{BF}^{d+1}Z_P^{d+1}$. In D dimensions the 
partition function for the BF model becomes~\cite{top}
\begin{eqnarray}
Z_{BF}^D=\left\lbrace
\begin{array}{ll}
T(M_D)^{-1} & \mbox{for D odd} \\
T(M_D)^{3-D \over D} & \mbox{for D even}  
\end{array}
\right .
\;,
\end{eqnarray}
where $T(M_D)$ is the Ray-Singer analitical torsion of the base manifold, and 
the integration over zero modes remains to be done.
\item On cohomologically trivial base manifolds both free models are 
identical, and it can be said on general grounds that the BF solutions label 
Proca formulations on sectors of the manifold with trivial structure.
\item There is a master action that connects both models. It is
\end{itemize}
\begin{equation}
S_{M}^{d+1}=\int_{M_{d+1}}\Bigl[{1 \over 8}b_{d-1}\wedge{}^{\star}b_{d-1}+
{\mu^2 \over 2}a \wedge{}^{\star}a-{1 \over 4}a \wedge H + 
{1 \over 2}(b_{d-1}-B_{d-1})\wedge F\Bigl]
\;.
\end{equation}
\vskip 0.5 truecm
\noindent{\bf Acknowledgements}

PJA would like to thank Alvaro Restuccia for very stimulating discussions.


\begin{thebibliography}{99}
%
\bibitem {rodolfo} X.~Fustero, R.~Gambini and A.~Trias, Phys. Rev. Lett.
{\bf 62} (1989) 1964;\\
R.~Gambini, Phys. Lett. {\bf B242} (1990) 398.
\bibitem {bala} 
C.~Aneziris, A.~P.~Balachandran, L.~Kauffman and A.~M.~Srivastava
, Int. J. Mod. Phys. {\bf A6} (1991) 2519;\\ 
A.~P.~Balachandran and P.~Teotonio-Sobrinho, Int. J. Mod. Phys. {\bf A9} 
(1994) 1569.
\bibitem {frac} G.~W.~Semenoff, Phys. Rev. Lett. {\bf 48} (1988) 517;\\ 
R.~Mackenzie and F.~Wilczek, Int. J. Mod. Phys. {\bf A3} (1988) 2827. 
\bibitem {sd} P.~K.~Townsend, K.~Pilch and P.~van Nieuwenhuizen, Phys. Lett. 
{\bf B136} (1984) 38;\\ C.~R.~Hagen, Ann. Phys. {\bf 157} (1984) 371.
\bibitem {schon} J.~Schonfeld, Nucl. Phys. {\bf B185} (1981) 157.
\bibitem {djt} S.~Deser, R.~Jackiw and S.~Tempelton, Phys. Rev. Lett. 
{\bf 48} (1982) 975; Ann. Phys. {\bf 140} (1982) 372; (E) {\bf 185} (1988) 406.
\bibitem {dj} S.~Deser and R.~Jackiw, Phys. Lett. {\bf B139} (1984) 371. 
\bibitem {grs} R.~Gianvittorio, A.~Restuccia and J.~Stephany, Mod. Phys. Lett. 
{\bf A6} (1991) 2121. 
\bibitem {ar} P.~J.~Arias and A.~Restuccia, Phys. Lett. {\bf B347} (1995) 241.
\bibitem {ar2} P.~J.~Arias, L.~Leal and A.~Restuccia, Phys. Lett. {\bf B367} 
(1996) 170.
\bibitem {stephany} J.~Stephany, Phys. Lett. {\bf B390} (1997) 128.
\bibitem {bos} R.~Banerjee, Nucl. Phys. {\bf B465} (1996) 157;\\
N.~Banerjee and R.~Banerjee, Mod. Phys. Lett. {\bf A11} (1996) 1919.
\bibitem {cremer} E.~Cremmer and J.~Scherk, Nucl. Phys. {\bf B72} (1974) 117. 
\bibitem {hagen} C.~R.~Hagen, Phys. Rev. {\bf D19} (1979) 2367.  
\bibitem {aurilia} A.~Aurilia and Y.~Takahashi, Prog. Theo. Phys. {\bf 66} 
(1981) 693.
\bibitem {dts} S.~Deser, P.~K.~Townsend and W.~Siegel, Nucl. Phys. {\bf B184} 
(1981) 333.
\bibitem {lahiri} T.~J.~Allen, M.~J.~Bowick and A.~Lahiri, Mod. Phys. Lett. 
{\bf A6} (1991) 559.
\bibitem {ft} D.~Z.~Freedman and P.~K.~Townsend, Nucl. Phys. {\bf B177} (1981) 
282.
\bibitem {town} P.~K.~Townsend, in {\it ``Gauge field theories: Theoretical   
studies and computer simulations''}, Proceedings of Karpacs Winter School, 
Ed. W.~Garczynski, Harwood Academic (1986), pp. 649.
\bibitem {adel} A.~Khoudeir, Mod. Phys. Lett. {\bf A11} (1996) 2489.
\bibitem {op} V.~I.~Ogievetski\u \i ~and V.~Polubarinov, Sov. J. Nuc. Phys. 
{\bf 4} (1967) 156. 
\bibitem {tp} Y.~Takahashi and R.~Palmer, Phys. Rev. {\bf D1} (1970) 
2974. 
\bibitem {kr} M.~Kalb and P.~Ramond, Phys. Rev. {\bf D9} (1974) 2273.
\bibitem {biz} C.~Bizdadea and S.~O.~Saliu, Nucl. Phys. {\bf B456} (1995) 473.
\bibitem {sen} P.~Senjanovic, Ann. Phys. {\bf 100} (1976) 227.
\bibitem {am} M.~Caicedo anb A.~Restuccia, Class. Q. Grav. {\bf 10} (1993) 
833.
\bibitem{tor} D.~B.~Ray and I.~M.~Singer, Advan. Math. {\bf 7} (1971) 145;\\
A.~Schwarz, Commun. Math. Phys. {\bf 67} (1979) 1.
\bibitem {top} M.~Blau and G.~Thompson, Ann. Phys. {\bf 205} (1991) 1.
\bibitem {hor} G.~T.~Horowitz, Commun. Math. Phys. {\bf 125} (1989) 417;\\
G.~Horowitz and M.~Srednicki, Commun. Math. Phys. {\bf 130} (1990) 83.
\bibitem {STy} A.Schwarz and Y.~Tyupkin, Nucl. Phys. {\bf B242} (1984) 436.

\end{thebibliography}
\end{document}